\documentstyle[12pt]{article}
\begin{document}

\baselineskip 24pt

\newcommand {\sheptitle}{Minimal Supersymmetric CPN Models}

\newcommand{\shepauthor}{K.J. Barnes}

\newcommand{\shepaddress}{Department of Physics \& Astronomy,\\
University of Southampton\\
Southampton, SO17 1BJ\\
United Kingdom\\ 
\smallskip
Telephone: +44 1703 592097\\
Email: kjb@phys.soton.ac.uk}

\newcommand{\shepabstract}
{\noindent {\bf Abstract} \hfill \\
\noindent Supersymmetric CPN models based on underlying bosonic 
Kahler
manifolds have not
been thought to arise directly from constrained linear ones.
A counterexample for
$N=4$ is presented using improved understanding of membranes
in superstring \mbox{theories}
leading to crucial central terms modifying the algebra
of supercharge densities.
The example has an immediate extension to all higher $N$. }

\begin{titlepage}
\vspace{.4in}
\begin{center}
{\large{\bf \sheptitle}}
\bigskip \\
\shepauthor
\\
{\shepaddress}
\vspace{.5in}
\end{center}
{\shepabstract}\\
\vfill
\end{titlepage}
It has recently been established that $CP2$ can be realised as a
non-linear 
supersymmetric model as the result of constraining a linear
supersymmetric model
\cite{1}.
Massless Goldstone bosons arise from components of global symmetries
which 
are spontaneously broken.  There is no extra symmetry for
Goldstone bosons in
supersymmetry.  Instead the supersymmetry forces complexification
of scalars.  This
leads to an increased number of massless excitations in
general, with complete
doubling of the original number in some cases.
Despite previously believed theorems
to the \mbox{contrary} by Lerche \cite{2} and Shore \cite{3},
the $CP2$ case was established as a
counter--example.
The key contribution leading to this possibility was that of Hughes 
and Polchinski \cite{4},
which showed that the original anticommutator for
supersymmetric charges had to be generalised to include a
central term at the 
underlying current density level.
This is a direct result of the more modern viewpoint
that supermembranes are just as fundamental as
elementary particles in string theory.
The key point seems to be that this is a case where the
symmetry of the hamiltonian 
is larger than the symmetry of the $S$-matrix.
When the anticommutator algebra for 
supersymmetic charges is generalised to local form as
\begin{equation}
\partial_\mu T \left( j_{A \alpha}^\mu (x) \bar{j}_{B 
\dot{\beta}}^\nu (y) \right) 
= 2 ( \sigma^ \rho)_{\alpha \dot{\beta}} T_\rho^\nu \delta^4 (x - y) 
\delta_{AB} 
+ 2 ( \sigma^\nu )_{\alpha \dot{\beta}} C_{AB} \delta^4 (x - y)
\end{equation}
the appearance of the central terms $C_{AB}$ is crucial.
The authors take advantage of the
fact that $T^{\mu \nu}$ is not the only unique conserved
symmetric tensor since $T^{\mu \nu} + C \eta^{\mu \nu}$ is also
conserved.
Thus equation (1) is clearly finite and Lorentz invariant, and from 
it
follow the usual consequences of degenerate multiplets for unbroken 
supersymmetries and Goldstone fermions for those that are broken.
In momentum
space, with $C_{AB}$ diagonal and $<T^{\mu \nu} > = \Lambda \eta^{\mu 
\nu}$,
this gives
\begin{equation}
q_\mu < j_{A \alpha}^\mu (q) \bar{j}_{A \dot{\beta}}^\nu > =
2 (\sigma^\nu)_{\alpha \dot{\beta}}
(\Lambda + C_{AA} ) + 0 (q)
\end{equation}
where there is no sum over $A$.
For those $A$ such that $\Lambda + C_{AA} \neq 0$, equation (2) 
implies a
$1/\! \not{\!q}$ singularity in the two current correlation;
$j_{A \alpha}^\mu$ couples the vacuum to a massless
fermion with coupling strength $[2 ( \Lambda + C_{AA} ) ]^{1/2}$,
where $\Lambda + C_{AA} \geq 0$.  It is now clear how to
evade the extra unwanted Goldstone bosons where the underlying coset
manifold is indeed Kahler.
The crucial point of extending the underlying algebra of supercharge
current densities by central terms has to be combined not merely
with a Kahler $G/H$,
but that manifold has to be reexpressed as a quotient of the
complexified $G$ (denoted as $G^C$) by a maximally extended
complexification of $H$ (denoted $\hat{H}$).
By
following the elegant treatment of Itoh, Kugor and Kunitoma
\cite{5}, this method will
display an explicit mapping manifesting the homeomorphism
between $G/H$ and $G^C/\hat{H}$. 
Since the bosonic coset space for $CP4$ is
\begin{equation}
\frac{G}{H} = \frac{SU_3}{SU_2 \times U_1} \;\; , \end{equation}
a convenient starting point for the appropriate notation
is given by the original Gell-
Mann matrices \cite{6}.
Note that $\lambda_8$ and $\lambda_3$ are in the
Cartan subalgebra, and that the
raising operators are $E_1 = 1/2 ( \lambda_1 + i \lambda_2 ) , E_2 = 
1/2 (\lambda_4 + i
\lambda_5 )$
and $E_3 = 1/2 (\lambda_6 + i \lambda_7),$
with
$E_{-1} = E_1^\dag$ , $E_{-2} = E_2^\dag$ and $E_{-3} = E_3^\dag$
as the lowering operators.
It is clear that all the raising and lowering operators are
nilpotent in this
representation.  This feature obviously extends to larger
$N$ and ensures that
constructing the Kahler potential is essentially
immediate in all cases.  Following 
reference \cite{5} a projection operator
$\eta$
with its only entry a one in the bottom right hand
corner is defined by
\begin{equation}
\eta = \frac{1}{3}  1 - \sqrt{ \frac{1}{3}} \lambda_8 \;\; ,
\end{equation}
and the complex subgroup $\hat{H}$ specified by the relationship
\begin{equation}
\hat{h} \eta = \eta \hat{h} \eta \;\; . \end{equation} 
This implies that the generators of
$\hat{H}$ are
$\lambda_8, \lambda_3, E_1, E_{-1}, E_{-2} $ and $ E_{-3}$,
and that
$E_{2}$ and $E_{3}$
are the four
elements of the algebra spanning $G^c / \hat{H}$.

Extending the notation of reference \cite{1},
the original (unconstrained) supersymmetric
action is constructed from nine (complex) chiral superfields.
In components, with
\begin{equation}
y^m = x^m + i \theta \sigma^m \bar{\theta} \;\; , \end{equation}
these have the form
\begin{eqnarray}
\Phi ( x, \theta , \bar{\theta} ) &=& \phi (y) + \sqrt{2} \theta 
\lambda_\phi (y) + \theta^2 F_{\phi}
(y) , \\
\Sigma_8 ( x, \theta, \bar{\theta} ) &=& \sigma_8 (y) + \sqrt{2} 
\theta \lambda_8 (y) + \theta^2
F_{8} (y) , \\
\Sigma_3 ( x, \theta, \bar{\theta} ) &=& \sigma_3 (y) + \sqrt{2} 
\theta \lambda_3 (y) + \theta^2
F_3 (y) , \\
\Delta_A (x, \theta, \bar{\theta} ) &=& \Delta_A (y) + \sqrt{2} 
\theta \Lambda_{A} (y) +
\theta^2 F_A (y) ,
\end{eqnarray}
where $A = (-1,1)$,
\begin{eqnarray}
\Delta_{-2} (x, \theta, \bar{\theta}) &=& \delta_{-2} (y) + \sqrt{2} 
\theta \Lambda_{-2} (y) +
\theta^2 F_{\Delta} (y) , \\
\Delta_{-3} (x, \theta, \bar{\theta} ) &=& \delta_{-3} (y) + \sqrt{2} 
\theta \Lambda_{-3} (y) +
\theta^2 F_{-3} (y), \\
\Gamma_\mu (x, \theta, \bar{\theta} ) &=& \gamma_\mu (y) + \sqrt{2} 
\theta \Omega_\mu (y)
+ \theta^2 F_\mu^\Gamma (y) ,
\end{eqnarray}
where $\mu = (2,3)$, $\sigma^m ( -1, \tau^a)$, and the $\tau^a$ are 
the Pauli matrices
$(a=1,2,3)$.  The chiral superfields 
transform under $SU_3$ as indicated by the index structure,
including $\Phi$ which is a
singlet.
The most general supersymmetric action is then written as
\begin{equation}
\begin{array}{c}
I = \int d^8 z \left[ \bar{\Phi} \Phi + \bar{\Sigma}_8 \Sigma_8 + 
\bar{\Sigma}_3 \Sigma_3
+ \bar{\Delta}_A \Delta_A + \bar{\Delta}_{-2} \Delta_{-2} + 
\bar{\Delta}_{-3} \Delta_{-3} 
+ \bar{\Gamma}_\mu \Gamma_\mu \right] \\
+ \int d^6 s W + \int d^6 \bar{s} \bar{W}
\end{array}
\end{equation}
where the superpotential $W$ is a functional of chiral superfields 
only.
Combining the 
eight non-singlet $SU_3$ superfields with their respective
matrices into the matrix
\begin{equation}
M = \Sigma_8 \lambda_8 + ... + \Gamma_\mu E_{-\mu} ,
\end{equation}
reveals that, under chiral $SU_3 \times SU_3$, $M$ transforms as
\begin{equation}
M \to L M R^\dag ,
\end{equation}
where the $\gamma_5$ structure is suppressed, and taking
\begin{equation} W = k \Phi \det M , \end{equation}
where $k$ is a constant, ensures that the model reduces to the
usual bosonic (Kahler) model below the symmetry breaking scale.
This starting action now yields
the potential
\begin{equation}
\begin{array}{c}
V = F_\phi \bar{F}_\phi + F_8 \bar{F}_8 + F_3 \bar{F}_3 + F_A 
\bar{F}_A +
F_{-2} \bar{F}_{-2} + F_{-3} \bar{F}_{-3} + F_\mu^\Gamma 
\bar{F}_\mu^\Gamma \\
= 4k^2 \phi \bar{\phi} \left[ \sigma_8 \bar{\sigma}_8 + \sigma_3 
\bar{\sigma_3} +
\delta_A \bar{\delta}_A + \delta_{-2} \bar{\delta}_{-2} + \delta_{-3} 
\bar{\delta}_{-3}
+ \gamma_\mu \bar{\gamma}_\mu \right] \\
+ k^2 \left[ \sigma_8^2 + \sigma_3^2 + \delta_A \delta_A +
\gamma_3 \delta_3 +
\gamma_\mu \gamma_{A+2} \right]
\left[ \bar{\sigma}_8^2 + \bar{\sigma}_3^2 + \bar{\delta}_A 
\bar{\delta}_A
+ \bar{\gamma}_{3} \bar{\delta}_{3} +
\bar{\gamma}_\mu \bar{\gamma}_{A+2} \right] 
\end{array} .
\end{equation}
In the formal limit as $k \to \infty$, the action becomes
\begin{equation}
I = \int d^8 z \frac{\Gamma_\mu \bar{\Gamma}_\mu}{4} ,
\end{equation}
as the constraints are satisfied by the superfield conditions
\begin{equation}
\Sigma_8 = \Sigma_3 = \Delta_A = \Delta_3 = \Gamma_3 = 0 . 
\end{equation}
The superfield $\Phi$ can again be ignored as a non-interacting 
spectator.
Notice that the
pair of complex superfields $\Gamma_\mu$ are all that remain in the
action, and they are not constrained.

In this notation the complex coset space is written in the form  
\cite{5}
\begin{equation}
L = \exp \left( \frac{- i}{2} \gamma_2 E_{2} \right)
exp \left( \frac{-i}{2} \gamma_3 E_3 \right) , \end{equation}
and this gives an explicit mapping of the homeomorphism between
$G/H$ and $G^c / \hat{H}$.
Following references \cite{5} the Kahler potential is given by
\begin{equation}
K = \ln {\det}_\eta \left[ exp
\left( \frac{-i \bar{\gamma_3} \bar{E}_3}{2} \right)
exp \left( \frac{-i \bar{\gamma_2} \bar{E}_2}{2} \right)
exp \left( \frac{-i \gamma_2 E_2}{2} \right)
exp \left( \frac{-i \gamma_3 E_3}{2} \right) \right]  ,
\end{equation}
where the notation indicates that the determinant is to be taken
in the bottom right hand of the matrix in this representation.
This reveals at once that
\begin{equation}
K = \ln \left[ 1 + \frac{\gamma_\mu \bar{\gamma}_\mu }{4}  \right],
\end{equation}
which is the desired result.
Notice how this presentation deals with the main objections which 
arose when it was 
claimed in reference \cite{1} that the generalisation directly to
CPN was possible.
It is not 
necessary to find special co-ordinates for the manifold in order
to demonstrate that it
is Kahler.
The $CPN$ manifolds are already known to be Kahler.
It is time that having
a general co-ordinate system in the $CP2$ case was very useful from a 
descriptive
viewpoint, but it is now clear that it was not really needed.
Of course it was very
convenient to use the nilpotency of $\tau^+$ and $\tau^-$
in the $CP2$ case, but far from being restricted
to that case it is now obvious that the number of nilpotent
matrices rises with $N$. 
Finally, there was the well established feature that there is an
increasing number of 
Kahler potentials with rising rank of $G$,
and each introduces an extra arbitrary 
constant.  Of course this current presentation
just gives one particular combination, 
but as is always the case with counterexamples one
is sufficient.

The author is grateful to Professor D.A.Ross for raising his
interest in this type of work.  This work is partly supported by 
PPARC
grant number GR/L56329.

\end{document}